\documentclass[12pt]{article}
\usepackage{amsmath,amsfonts,amssymb}
\usepackage[dvips]{color}

\begin{document}
\thispagestyle{empty}

\begin{titlepage}
\strut\hfill UMTG--267
\vspace{.5in}
\begin{center}

\centerline{\Large \bf Finite-size effect for four-loop Konishi}
\vskip .5cm
\centerline{\Large \bf of the $\beta$-deformed ${\cal N}=4$ SYM}

\end{center}
\vskip 1.2cm

\centerline{\bf Changrim  Ahn$^1$, Zoltan Bajnok$^2$, Diego
Bombardelli$^1$, and Rafael I. Nepomechie$^3$ }

\vskip 1.2cm

\centerline{\sl $^1$Department of Physics and Institute for the
Early Universe} \centerline{\sl Ewha Womans University, DaeHyun
11-1, Seoul 120-750, S. Korea}

\vskip 0.8cm

\centerline{\sl $^2$Theoretical Physics Research Group} \centerline{\sl
Hungarian Academy of Sciences, 1117 Budapest, P\'azm\'any s. 1/A
Hungary}

\vskip 0.8cm

\centerline{\sl $^3$Physics Department, P.O. Box 248046}
\centerline{\sl University of Miami, Coral Gables, FL 33124, U.S.A.}

\vskip 20mm

\baselineskip 18pt

\begin{center}
{\bf Abstract}
\end{center}
We propose that certain twists of the $su(2|2)$ $S$-matrix elements describe
the $\beta$-deformation of ${\cal N}=4$ supersymmetric Yang-Mills theory.
We compute the perturbative four-loop
anomalous dimension of the Konishi operator of the deformed gauge theory
from the L\"uscher formula based on these twisted $S$-matrix elements.
The result agrees exactly with the perturbative gauge theory computations.

\end{titlepage}
\newpage
\baselineskip 18pt

\def\nn{\nonumber}
\newcommand{\bea}{\begin{eqnarray}}
\newcommand{\eea}{\end{eqnarray}}
\newcommand{\be}{\begin{eqnarray}}
\newcommand{\ee}{\end{eqnarray}}
\newcommand{\ben}{\begin{eqnarray*}}
\newcommand{\een}{\end{eqnarray*}}
\newcommand{\beq}{\begin{equation}}
\newcommand{\eeq}{\end{equation}}

\def\bJ{{\mathbb J}}
\def\bR{{\mathbb R}}
\def\bL{{\mathbb L}}
\def\bQ{{\mathbb Q}}
\def\bC{{\mathbb C}}
\def\bH{{\mathbb H}}
\newcommand{\non}{\nonumber}
\newcommand{\id}{\mathbb{I}}
\newcommand{\tr}{\mathop{\rm tr}\nolimits}
\newcommand{\sgn}{\mathop{\rm sign}\nolimits}
\newcommand{\diag}{\mathop{\rm diag}\nolimits}
\newcommand{\cf}{\ensuremath{\mathfrak{c}}}

\vskip 0cm

\renewcommand{\thefootnote}{\arabic{footnote}}
\setcounter{footnote}{0}

\section{Introduction}
The AdS/CFT correspondence \cite{AdS/CFT} is a duality between the ${\cal N}=4$
super Yang-Mills (SYM) theory and type IIB string theory on $AdS_5\times S^5$.
Integrability in the planar limit \cite{MinZar} has played a crucial role in
clarifying the correspondence, which relates strong and weak couplings.

Asymptotic Bethe ansatz equations (BAEs) for all orders and sectors,
which have provided a computational tool, were conjectured in \cite{BeiSta}.
Another important approach was to construct an exact $S$-matrix based on the
integrability \cite{Staudacher}.
The $S$-matrix for fundamental excitations was determined from the $su(2|2)$
symmetry algebra in \cite{Beisert,AFZ} up to an overall scalar factor which was
fixed in \cite{BHL,BES} using crossing symmetry \cite{Janik}.
The conjectured BAEs can be derived by diagonalizing the Bethe-Yang matrix
\cite{Beisert,MarMel}.
These asymptotic BAEs, however, have a fundamental limitation since they are valid
only when the size of the spin chain becomes infinite.
It has been pointed out in \cite{AJK} that the asymptotic BAEs should fail because
of wrapping interactions which arise when the order of perturbation theory goes
beyond the size of the spin chain.

The deviations from the asymptotic BAEs have been quantitatively studied in the
weak-coupling limit using generalized L\"uscher corrections and compared with
perturbative SYM computations \cite{FSSZ0} in a series of
papers \cite{BajJan,BajJanLuk}.
The results show that the L\"uscher corrections agree exactly with the deviations
which is another triumph for the exact $S$-matrix of the ${\cal N}=4$ SYM.

Our main interest in this paper is to generalize the wrapping correction analysis
to the $\beta$-deformed SYM theory.
The deformed SYM theory is obtained by replacing the original
${\cal N}=4$ superpotential for the chiral superfields by:
\beq
W= ih\ {\rm tr}
(e^{i\pi\beta}\phi\psi Z-e^{-i\pi\beta}\phi Z\psi).
\eeq
The deformation breaks the supersymmetry down to ${\cal N}=1$ but still maintains
the conformal invariance in the planar limit to all perturbative orders
\cite{LeiStr, MPSZ}, since the deformation becomes exactly marginal
for real $\beta$ if
\beq
h{\overline h}=g_{\rm YM}^2,
\eeq
where $g_{\rm YM}$ is the Yang-Mills coupling constant.

Under the AdS/CFT duality, it is believed that this $\beta$-deformed SYM theory is
related to a string theory with the Lunin-Maldacena background \cite{LunMal}.
When the deformation parameter is real,
the string theory on this deformed background maintains
the classical integrability \cite{Frolov,BykFro}, and has identical excitations
such as giant magnons, whose finite-size effects have been obtained by transforming
the $AdS_5\times S^5$ background under some T-duality \cite{BykFro}.
Perturbative integrability for the deformed SYM was studied in
\cite{FrRoTs,BerChe, BeiRoi}.
An important development in the deformed SYM theory was the
perturbative computation in \cite{FSSZ,FSSZ1} of anomalous dimensions for
the one and two magnon states in the $su(2)$ sector up to four loops.
These hints of the integrability of the deformed SYM theory can be established firmly
if one can construct an exact factorizable $S$-matrix which is consistent with the
wrapping corrections.

In this note we propose that certain twists of the $su(2|2)$ $S$-matrix
elements describe the $\beta$-deformed SYM theory.
We show that, via the L\"uscher formula, these twisted amplitudes lead to
the correct wrapping corrections for the $su(2)$ Konishi operator.


\section{Finite-size effect of the $\beta$-deformed SYM}

\subsection{Asymptotic Bethe ansatz equation}

The asymptotic BAEs have been conjectured for the $\beta$-deformed
SYM theory in \cite{FrRoTs,BeiRoi}.
The BAEs become quite simple for the $su(2)$ sector which is relevant to the Konishi
operator.
There are two ``impurities'' which carry two Bethe roots $u_1, u_2$ in the
spin chain of length $L=4$ which satisfy
\beq
e^{ip_1 L}=e^{2\pi i\beta L}e^{2i\theta(p_1,p_2)}\frac{u_1-u_2+i}{u_1-u_2-i},\qquad
e^{ip_2 L}=e^{2\pi i\beta L}e^{-2i\theta(p_1,p_2)}\frac{u_2-u_1+i}{u_2-u_1-i}.
\eeq
The momenta $p_i$ are related to the Bethe roots by
\beq
u_i=\frac{1}{2}\cot\frac{p_i}{2}\sqrt{1+16g^2\sin^2\frac{p_i}{2}},
\eeq
where
\beq
g^2=\frac{g_{\rm YM}^2 N}{16\pi^2}.
\eeq
The energy for a magnon with momentum $p_i$ is given by
\beq
E(p_i)=\sqrt{1+16 g^2\sin^2\frac{p_i}{2}}.
\eeq

One can deduce the momentum conservation relation from the BAEs
\beq
p_1+p_2=4\pi\beta.\label{momcons}
\eeq
We will define shifted momentum variables by
\beq
{\tilde p}_i\equiv p_i-2\pi\beta,\qquad {\tilde p}_1+{\tilde p}_2=0.
\eeq
The BAE and the energy are expressed simply in terms of ${\tilde p}={\tilde p}_1$
as follows:
\be
e^{4i{\tilde p}}&=&e^{2i\theta({\tilde p}+2\pi\beta,-{\tilde p}+2\pi\beta)}
\left(\frac{u({\tilde p}+2\pi\beta)-u(-{\tilde p}+2\pi\beta)+i}
{u({\tilde p}+2\pi\beta)-u(-{\tilde p}+2\pi\beta)-i}\right)\\
E_{\rm total}(\beta)&=&E({\tilde p}+2\pi\beta)+E(-{\tilde p}+2\pi\beta).
\ee

We can solve the BAE perturbatively in $g^2$ by
expanding both ${\tilde p}$ and $E$
\begin{eqnarray}
{\tilde p}&=&{\tilde p}_0+g^2{\tilde p}^{(1)}+g^4{\tilde p}^{(2)}+g^6{\tilde p}^{(3)}
+\ldots\\
E_{\rm total}(\beta)&=&2+g^2 E_1(\beta)+g^4 E_2(\beta)+g^6 E_3(\beta)+g^8 E_4(\beta)
+\ldots.
\end{eqnarray}
The leading order BAE is
\beq
e^{4 i\tilde{p}_0} =
\frac{\cot (\frac{\tilde{p}_0}{2} - \pi\beta)+
\cot (\frac{\tilde{p}_0}{2} + \pi\beta)+2i}{
\cot (\frac{\tilde{p}_0}{2} - \pi\beta)+
\cot (\frac{\tilde{p}_0}{2} + \pi\beta)-2i}
\eeq
with a solution
\beq
\cos\tilde{p}_0=\frac{1\mp 3\Delta}{4\cos (2\pi\beta)},
\label{leading}
\eeq
where we defined $\Delta$ by \cite{FSSZ}
\beq
\qquad \Delta\equiv \frac{\sqrt{5+4\cos(4\pi\beta)}}{3}.
\eeq
From now on we focus on the ``$-$'' sign only keeping in mind that the other solution
can be obtained by changing the sign of $\Delta$.
The energy in the leading order becomes
\beq
E_1(\beta)=8\sin^2\left(\frac{\tilde{p}_0}{2}-\pi\beta\right)
+8\sin^2\left(\frac{\tilde{p}_0}{2}+\pi\beta\right)
=6(1+\Delta)\label{baei}.
\eeq

Higher order BAEs and their solutions can be obtained iteratively along with
the energy corrections as follows:
\be
E_2&=&-\frac{3}{\Delta}-15-21\Delta-9\Delta^2\label{baeii}\\
E_3&=&-\frac{3}{4\Delta^3}+\frac{153}{4\Delta}+114+\frac{495}{4}\Delta+54\Delta^2
+\frac{27}{4}\Delta^3 \label{baeiii}\\
E_4&=&\frac{3\,{\left( 1 + \Delta  \right) }^4(-1 - 2\Delta  + 49{\Delta }^2
+ 84{\Delta }^3 - 1359{\Delta }^4 - 5562{\Delta }^5 - 2673{\Delta }^6
+ 1944{\Delta}^7)}{8\Delta^5(1+3\Delta)^2}\nonumber\\
&+&\left(-\frac{9}{\Delta}+27+54\Delta-90\Delta^2-189\Delta^3-81\Delta^4\right)\zeta(3),
\label{baeivii}
\ee
where the term proportional to $\zeta(3)$ originates from the
dressing phase \cite{BES}.

\subsection{Perturbative gauge theory results}

We summarize here the perturbative computation of anomalous dimensions of the $su(2)$
Konishi operators ${\rm Tr}(XZXZ)$ and ${\rm Tr}(ZZXX)$ of the $\beta$-deformed SYM
up to four loops \cite{FSSZ}.
One of the two eigenvalues of the dilatation operator is given by
\be
\gamma&=&4+g^2 \gamma_1+g^4 \gamma_2+g^6 \gamma_3+g^8 \gamma_4+\ldots\non\\
\gamma_1&=&6(1+\Delta)\label{siegi}\\
\gamma_2&=&-\frac{3}{\Delta}-15-21\Delta-9\Delta^2\label{siegii}\\
\gamma_3&=&-\frac{3}{4\Delta^3}+\frac{153}{4\Delta}+114+\frac{495}{4}
\Delta+54\Delta^2+\frac{27}{4}\Delta^3
\label{siegiii}\\
\gamma_4&=&-\frac{3}{8\Delta^5}+\frac{33}{2\Delta^3}-\frac{1701}{4\Delta
}-1230-\frac{2427}{2}\Delta
-180\Delta^2+162\Delta^4+\frac{2997}{8}\Delta^3\non\\
&+&\left(-\frac{9}{\Delta}+297+702\Delta+234\Delta^2-405\Delta^3-243\Delta^4\right)
\zeta(3)-360\left(1+\Delta\right)^2\zeta(5)\label{siegiviii},
\ee
and the other eigenvalue is obtained by changing the sign of $\Delta$.

Up to $g^6$ order these results match exactly with the asymptotic BAE results.
One can compare (\ref{siegi}) with (\ref{baei}), (\ref{siegii}) with (\ref{baeii}),
and (\ref{siegiii}) with (\ref{baeiii}).
At wrapping order $g^8$ there is a discrepancy between (\ref{siegiviii}) and
(\ref{baeivii}),
\beq
\Delta E_{\rm wrapping}
=g^8\left[-54(1+\Delta)^3(-5+3\Delta)\zeta(3)-360(1+\Delta)^2\zeta(5)+
\frac{81\,{\left( 1 - 3\,\Delta  \right) }^2\,{\left( 1 + \Delta  \right) }^4}
{{\left( 1 + 3\,\Delta  \right) }^2}
\right].\label{differ}
\eeq
We will explain this difference at the leading wrapping order by a finite-size
effect based on the L\"uscher formula.

\subsection{L\"uscher formula}

The L\"uscher formula for multi-particle states for a theory with non-diagonal
$S$-matrix has been proposed and checked in \cite{BajJan}.
For the case of the Konishi operator, the wrapping correction is given by
\beq
\Delta E_{\rm wrapping}=-\sum_{\ell=1}^{\infty}\int_{-\infty}^{\infty}\frac{dq}{2\pi}
\left(\frac{z^-}{z^+}\right)^{L}\sum_{j,j'} (-1)^{F_{(jj')}}\left[
{\cal S}^{(\ell 1)}(z^{\pm},x_1^{\pm})
{\cal S}^{(\ell 1)}(z^{\pm},x_2^{\pm})\right]_{(jj')(11)}^{(jj')(11)},
\label{luscher}
\eeq
where the rapidities of physical particles are given by
\beq
x_i^{\pm}=\frac{1}{4g}\left(\cot\frac{p_i}{2}\pm i\right)
\left(1+\sqrt{1+16g^2\sin^2\frac{p_i}{2}}\right)
\eeq
and that of the mirror $\ell$-particle bound states is
\beq
z^{\pm}=\frac{q+i\ell}{4g}\left(\sqrt{1+\frac{16g^2}{\ell^2+q^2}}\pm 1\right).
\eeq
$F_{(jj')}=F_j+F_{j'}$ is given by
\beq
F_j=\begin{cases}
0&\text{if}\quad j=1,\ldots,2\ell\\
1&\text{if}\quad j=2\ell+1,\ldots,4\ell.
\end{cases}
\eeq
As explained in \cite{BajJan}, we need to choose $L=4$ because we are including the
`string frame' phase factors into the effective length.

The total $S$-matrix is a tensor product of two twisted $su(2|2)$ $S$-matrices,
\beq
{\cal S}^{(\ell 1)}(z^{\pm},x_i^{\pm})=
S^{(\ell 1)}_{\rm scalar}(q,u_i)\left[
{\cal S}^{(\ell 1)}_{\rm matrix}(q,u_i)\otimes {\cal S}^{(\ell 1)}_{\rm matrix}(q,u_i)\right],
\eeq
and the $S$-matrix part in the L\"uscher formula can be rewritten as
\be
&&\sum_{j,j'} (-1)^{F_{(jj')}}\left[
{\cal S}^{(\ell 1)}(q,u_1)
{\cal S}^{(\ell 1)}(q,u_2)\right]_{(jj')(11)}^{(jj')(11)}\nonumber\\
&=&S^{(\ell 1)}_{\rm scalar}(q,u_1)S^{(\ell 1)}_{\rm scalar}(q,u_2)
\left(\sum_{j} (-1)^{F_{j}}\left[{\cal S}^{(\ell 1)}_{\rm matrix}(q,u_1)
{\cal S}^{(\ell 1)}_{\rm matrix}(q,u_2)\right]_{j1}^{j1}\right)^2.
\ee

For the $su(2)$ Konishi operator it is enough to consider scattering amplitudes
between $\ell$-particle bound states and $A_1$ which are diagonal.
We propose that these matrix elements are given explicitly by
\begin{equation}
{{\cal S}^{(\ell 1)}}_{j1}^{j1}=
\begin{cases}
{S^{(\ell 1)}}_{j1}^{j1}=a^5_5,&\quad j=1,\ldots,\ell+1,\\
{S^{(\ell 1)}}_{j1}^{j1}=2 a^8_8,&\quad j=\ell+2,\ldots,2\ell
\end{cases}
\label{bossca}
\end{equation}
for the bosonic states and
\beq
{{\cal S}^{(\ell 1)}}_{j1}^{j1}=
\begin{cases}
e^{i\pi\beta}{S^{(\ell 1)}}_{j1}^{j1}=
e^{i\pi\beta}a^9_9,&\quad j=2\ell+1,\ldots,3\ell,\\
e^{-i\pi\beta}{S^{(\ell 1)}}_{j1}^{j1}
=e^{-i\pi\beta}\frac{a^9_9+a^3_3}{2},&\quad j=3\ell+1,\ldots,4\ell
\end{cases}
\label{fersca}
\eeq
for the fermionic states.
Explicit expressions for the undeformed matrix elements $a_{j}^{j}$
are given in \cite{BajJan,BajJanLuk}.

Let us emphasize that contrary to \cite{BajJan} we calculate here the L\"uscher
correction of the $su(2)$ representative of the Konishi operator.
In the $\beta$-deformed theory the two representatives have different anomalous
dimensions and the direct comparison is available only in the $su(2)$ case.
This calculation is also new in the undeformed case where the L\"uscher correction
for only the $sl(2)$ representative has been calculated so far.
There are calculations in the $su(2)$ sector based on the $Y$-system \cite{GKV}
but we are not able to use this approach since the deformed-TBA equations
are not available.

Since the exponential factor becomes
\beq
\left(\frac{z^-}{z^+}\right)^4=\frac{256g^8}{(q^2+\ell^2)^4}+\ldots,
\eeq
we may consider only a leading term in each expression.
The scalar part of the $S$-matrix is the same as the undeformed case,
\beq
S^{(\ell 1)}_{\rm scalar}(q,u)=
\frac{[q - i(\ell - 1) -2 u]}{[q + i(\ell- 1) - 2 u]}
\frac{ (2 u +i)^2}{[(q-2u)^2+(\ell+ 1)^2]}.
\eeq
The sum in the matrix part is nontrivial.
Taking the $g\to 0$ limits on the $S$-matrix elements given in Eqs.(\ref{bossca})
and (\ref{fersca}),
we obtain at leading order
\beq
{\cal S}^{(\ell 1)}_{\rm matrix}(q,u)_{j1}^{j1}
\approx
\begin{cases}
\frac{2u-q-i(\ell-1)}{2u+i},&\quad
j=1,\ldots,\ell+1,\\
\frac{(2u-q)^2+(\ell+1)^2}{(2u+i)(2u-q+i(\ell-1))},&\quad
j=\ell+2,\ldots,2\ell,\\
e^{i\pi\beta}\frac{2u-q-i(\ell+1)}{\sqrt{4u^2+1}},&\quad
j=2\ell+1,\ldots,3\ell,\\
e^{-i\pi\beta}\frac{(2u-i)(\ell^2+(2u-q+i)^2)}{(2u+i)(2u-q+i(\ell-1))\sqrt{4u^2+1}},
&\quad j=3\ell+1,\ldots,4\ell.
\end{cases}
\eeq
Inserting these into Eq.(\ref{luscher}) and integrating using the residue at $q=i\ell$,
we get \footnote{Summing up the contributions of the other residues gives a vanishing result.}
\beq
\Delta E_{\rm wrapping}=\sum_{\ell=1}^{\infty}\left[
\frac{f_1(u_1,u_2)}{\ell^5}+\frac{f_2(u_1,u_2)}{\ell^3}+f_3(u_1,u_2,\ell)\right]
\label{wrapping}
\eeq
where\footnote{
We have expressed the deformation parameter $\beta$ in terms of
$u_1,\ u_2$ using Eq.(\ref{momcons}) for simplicity.}

\beq
f_1(u_1,u_2)=-\frac{2560(1+2u_1^2+2u_2^2)^2}{(4u_1^2+1)^2(4u_2^2+1)^2}
\label{fone}
\eeq
and
\be
f_2&=&\frac{\rm num}{(4u_1^2+1)^4(4u_2^2+1)^4}\\
{\rm num}&=&2048\left(-1+5u_1^2+48u_1^4+96u_1^6-2u_1u_2-16u_1^3u_2-32u_1^5u_2
+5u_2^2+224u_1^2u_2^2\right.\nonumber\\
&+&1024u_1^4u_2^2+1536u_1^6u_2^2+768u_1^8u_2^2-16u_1u_2^3
-128u_1^3u_2^3+64 u_1^8\nonumber\\
&-&256u_1^5u_2^3+48u_2^4+1024u_1^2u_2^4+3200u_1^4u_2^4
+2560u_1^6u_2^4-32u_1u_2^5\nonumber\\
&-&\left.256u_1^3u_2^5-512u_1^5u_2^5+96u_2^6
+1536u_1^2u_2^6+2560u_1^4u_2^6+64u_2^8+768u_1^2u_2^8\right).
\ee
The $f_3$ is too complicated to write (1775 terms in the numerator).

The BAE roots $u_1,u_2$ at the leading order can be used to evaluate the wrapping
correction,
\beq
u_1=\frac{1}{2}\cot\left(\frac{{\tilde p}_0}{2}+\pi\beta\right),\qquad
u_2=\frac{1}{2}\cot\left(-\frac{{\tilde p}_0}{2}+\pi\beta\right).
\eeq
Using ${\tilde p}_0$ in Eq.(\ref{leading}), we obtain
\bea
u_1&=&\frac{{\left( 1 - 3\,\Delta  \right) }^2}
  {2\,{\sqrt{-1 + 9\,{\Delta }^2}}\,
    \left( 3\,{\sqrt{1 - {\Delta }^2}} +
      2\,{\sqrt{1 + \frac{2}{1 + 3\,\Delta }}}
      \right) },\\
u_2&=&\frac{{\left( 1 - 3\,\Delta  \right) }^2}
  {2\,{\sqrt{-1 + 9\,{\Delta }^2}}\,
    \left( 3\,{\sqrt{1 - {\Delta }^2}} -
      2\,{\sqrt{1 + \frac{2}{1 + 3\,\Delta }}}
      \right) }.
\eea
Inserting these exact expressions into Eq.(\ref{wrapping}) and
summing up the infinite terms exactly, we obtain
\beq
\Delta E_{\rm wrapping}=g^8\left[-54(1+\Delta)^3(-5+3\Delta)\zeta(3)
-360(1+\Delta)^2\zeta(5)+
\frac{81\,{\left( 1 - 3\,\Delta  \right) }^2\,{\left( 1 + \Delta  \right) }^4}
{{\left( 1 + 3\,\Delta  \right) }^2}
\right]
\eeq
which matches exactly with Eq.(\ref{differ}).
The wrapping correction for the second anomalous dimension can be computed from
the same L\"uscher formula by simply replacing $\Delta$ with $-\Delta$.

The proposed $S$-matrix elements can be constructed from a Drinfeld twist of the
$su(2|2)$ $S$-matrix \cite{drinfeld}
which we will report in a separate publication \cite{future}.
Furthermore this $S$-matrix can lead to the conjectured asymptotic BAEs via
nested BAE analysis.
We hope this will lead to studying the $\beta$-deformed SYM theory in a
more rigorous way.

There are several imminent questions.
While our analysis successfully generated the four-loop L\"uscher corrections of
the $su(2)$ Konishi operator, we have not succeeded in the one impurity system, whose
perturbative results are given in \cite{FSSZ,FSSZ1}, and for which L\"uscher corrections
have been calculated for $\beta=1/2$ and generic values of $L$ in \cite{single}.
Another challenge is to derive the asymptotic BAEs based on our $S$-matrices which will
eventually lead to the thermodynamic Bethe ansatz, thereby 
generalizing the TBA equations in \cite{TBA}.
We hope to address these issues in the near future.

\vspace{.2in}

{\em Note Added:} 
In the recent paper \cite{GLM}, our result for the four-loop L\"uscher
correction of the $su(2)$ Konishi operator is confirmed, and also
results for a single impurity are obtained using a (non-symmetric)
deformation of the $Y$-system.

\section*{Acknowledgements}
We thank N. Beisert, M. Martins, and
M. Staudacher for valuable discussions and the APCTP Focus Program
2009 and 2010 where parts of the work have been performed.
This work was supported in part by KRF-2007-313- C00150, WCU Grant
No. R32-2008-000-101300 (CA, DB), by Bolyai Scholarship and OTKA 81461 (ZB),
and by the National Science Foundation under Grants
PHY-0554821 and PHY-0854366 (RN).

\end{document}